\documentclass[12pt]{article}

\newcommand{\be}{\begin{equation}}
\newcommand{\ee}{\end{equation}}
\newcommand{\bea}{\begin{eqnarray}}
\newcommand{\eea}{\end{eqnarray}}

\newcommand{\vek}{\vec{k}}

\newcommand{\half}{\frac{1}{2}}

\newcommand{\rhat}{\hat{r}}

\newcommand{\that}{\hat{t}}

\newcommand{\udot}{\dot{u}}
\newcommand{\vdot}{\dot{v}}
\newcommand{\Udot}{\dot{U}}
\newcommand{\Vdot}{\dot{V}}
\newcommand{\Rdot}{\dot{R}}

\begin{document}

\begin{titlepage}
\begin{flushleft}
       \hfill                      {\tt gr-qc/0005115}\\
       \hfill                      HIP-2000-25/TH \\
       \hfill                      June 21, 2000\\
\end{flushleft}
\vspace*{3mm}
\begin{center}
{\Large {\bf Hawking Radiation from AdS Black Holes \\}}
\vspace*{12mm}
{\large
Samuli Hemming\footnote{E-mail: samuli.hemming@hip.fi} \\
Esko Keski-Vakkuri\footnote{E-mail: esko.keski-vakkuri@hip.fi}\\}

\vspace{5mm}

{\em Helsinki Institute of Physics \\
P.O. Box 9\\
FIN-00014  University of Helsinki \\
Finland }
\vspace*{10mm}
\end{center}


\begin{abstract}
We investigate Hawking radiation from black holes
in (d+1)-dimensional anti-de Sitter space. We focus on s-waves,
make use of the geometrical optics approximation, and follow three
approaches to analyze the radiation.
First, we compute a Bogoliubov transformation between
Kruskal and asymptotic coordinates and compare the different
vacua. Second, following a method due to Kraus, Parikh, and
Wilczek, we view Hawking radiation as a tunneling process across
the horizon and compute the tunneling probability. This approach
uses an anti-de Sitter version of a metric originally introduced
by Painlev\'{e} for Schwarzschild black holes. From the tunneling
probability one also finds a leading correction to the
semi-classical emission rate arising from backreaction to the
background geometry. Finally, we consider a spherically symmetric
collapse geometry and the Bogoliubov transformation
between the initial vacuum state and the vacuum of an asymptotic
observer.

\end{abstract}

\end{titlepage}

\baselineskip16pt

\section{Introduction}

Anti-de Sitter black holes play a major role in the AdS/CFT conjecture
\cite{Maldacena:1997re} (see \cite{Aharony:1999ti} for a review),
and they have also
received interest in the context of brane-world scenarios
based on the setup of Randall and Sundrum \cite{Randall:1999ee,
Randall:1999vf}.
The purpose of this note is to briefly investigate the basic property
of black holes, Hawking radiation, in anti-de Sitter spacetimes.

In the 2+1 dimensional special case, the BTZ black holes
\cite{Banados:1992wn}, Hawking radiation of a massles conformally
coupled scalar field was investigated in detail by Hyun {\em et.
al.} \cite{Hyun:1994na}. Here we will instead take a more generic
approach (while sacrifying detail), and extend some standard ways
to analyze Hawking radiation from Schwarzschild black holes to the
anti-de Sitter case in a generic dimension. The standard methods
in question are to investigate eternal black holes and the role of
different vacua based on different boundary conditions at the
horizons, and to investigate radiation in a spherical collapse
geometry while the quantum field is in a natural initial vacuum.
Customarily, the focus is on finding the leading thermal
characteristics of the radiation. For this purpose it is
sufficient to invoke a geometric optics approximation which
neglects backscattering of outgoing waves from the spacetime
curvature and essentially reduces the problem to a two dimensions.
The neglected effects would give rise to a gray-body factor in the
thermal emission spectrum. The gray-body factors can be found by
investigating absorption by AdS black holes.

We will also investigate the back-reaction to the geometry
by taking into account the selfinteraction effect analyzed by
Kraus, Parikh and Wilczek \cite{Kraus:1994fh, Kraus:1995by,
Parikh:1999mf} (see also \cite{Keski-Vakkuri:1997xp}).
This approach views Hawking radiation
as a tunneling process across the horizon. Here we extend the approach
of \cite{Kraus:1994fh, Kraus:1995by,
Parikh:1999mf} to the anti-de Sitter case, and find similar results.

\section{Hawking radiation from eternal black holes}

One standard method to derive the Hawking radiation is to use a
Bogoliubov transformation between two basis of annihilation
and creation operators corresponding to mode expansions
of the field operator in two preferred coordinate systems
used to describe a black hole: the asymptotic coordinates and the
Kruskal coordinates. In the asymptotic coordinates,
the AdS$_{d+1}$ black hole metric is (see {\em e.g.} \cite{Horowitz:1999jd})
\be
ds^2 = -F(r) \,dt^2 + \frac {dr^2}{F(r)} + r^2 d\Omega_{d-1}^2
\label{adsbh}
\ee
where
\be
 F(r) = 1- \frac{\mu}{r^{d-2}}+r^2 \ ,
\ee
and we work in units where the AdS radius $l=1$.
The parameter $\mu$ is proportional to the ADM mass $M$ of the black hole,
\be
 M = \frac{(d-1)A_{d-1}}{16\pi G_{d+1}} \mu
\ee
where $G_{d+1}$ is Newton's constant in $d+1$ dimensions and
$A_{d-1}=2\pi^{d/2}/\Gamma (d/2)$ is the volume of
a unit $(d-1)$-sphere. The explicit formula for the horizon
radius $r_H$
can be found by solving the polynomial equation
$F(r_H)=0$. For example,
in $4+1$ dimensions,
\be
r_H = \sqrt{ \frac 12 \left( \sqrt{ 1+ 4\mu } -1 \right)} \ .
\ee
Then, the Hawking temperature $T_H$ can be found by looking at the
periodicity of the Euclidian section of the metric near $r_H$. The
generic result is
\be
    T_H = \frac{1}{4\pi} F'(r_H)
\ee
where $'=d/dr$. E.g. in 4+1 dimensions,
\be
T_H = \frac {\sqrt{1 + 4\mu}}{2 \pi  r_H} \ .
\ee

To investigate mode solutions of a field equation, it
is useful to introduce the tortoise coordinate $r_*$,
\be
r_* = \int^r_{r_H} \frac{d\rhat}{F(\rhat )} \ .
\ee
Again, in $4+1$ dimensions the explicit formula is
\be
r_* = \frac{1}{\sqrt{1+4\mu}} \left( r_0 ~{\rm arctan} \left(
\frac {r}{r_0} \right) + \frac 12 r_H ~{\rm ln} \left( \frac
{r-r_H}{r+r_H} \right) \right) \ ,
\ee
where $r_H$ is the horizon radius
and $r_0$ is a shorthand notation denoting a radius
\be
r_0 =  \sqrt{ \frac 12 \left( \sqrt{ 1+ 4\mu } +1 \right)} \ .
\ee
(In 4+1 dimensions, the
equation $F(r)=0$ has 4 complex zeroes: 2 real zeroes at $\pm r_H$
and 2 imaginary zeroes at $\pm ir_0$.)
It is also convenient to introduce the null coordinates
\bea
u &=& t - r_* \nonumber \\
v &=& t + r_*
\eea
where $r_*$ is the tortoise coordinate. Then the metric takes the
form
\be
 ds^2 = -F(r) \,du \,dv + r^2 d\Omega_{d-1}^2
\label{metric}
\ee
and the solutions to the wave equation are infalling and outgoing
partial waves.

In the region outside the past and future horizons, the
Kruskal coordinates $U, ~V$ are defined as follows:
\bea
U &=& -{\rm exp} \left( - 2\pi T_H~u \right)
\nonumber \\
V &=& {\rm exp} \left( 2\pi T_H ~v \right)
\eea
In $U,~V$ coodinates, the metric takes the form
\be
ds^2 = -\frac{F(r)e^{-4\pi T_H r_*}}{(2\pi T_H)^2}dUdV
+ r^2 d\Omega^2_{d-1} \ .
\ee
In 4+1 dimensions, the explicit form is
\be
ds^2 = - \frac {r_H^2}{1 + 4\mu} (1 + \frac {r_0^2}{r^2})
(r_H + r)^2 ~{\rm exp} \left( - \frac {2 r_0}{r_H}
~{\rm arctan} \left( \frac {r}{r_0} \right) \right) \,dU\,dV
+ r^2 d\Omega^2_3 \ .
\ee
In Kruskal coordinates, the metric
can be extended over the whole spacetime, except for the
origin $r=0$, which corresponds
to the true curvature singularity of the black hole.

Next, we will focus on s-waves, adopt the geometric optics
approximation, and truncate to two dimensions, following
the classic paper by Unruh \cite{Unruh:1976db}. The discussion
also overlaps with \cite{Klemm:1998bb}\footnote{We thank D. Klemm
for bringing this reference to our attention.}.

For a quantum field in the black hole background, there are two
canonical choices for a natural vacuum state. If one wants to
mimic a situation where the black hole is created by collapsing
matter, one requires the field to be in a vacuum corresponding
absence of positive energy modes in the $U$ and $v$ coordinates
near the past horizon $V=0$. This boundary condition refers only
to the past of the asymptotic region of spacetime, an is known as
the Unruh vacuum. Another vacuum choice, the Hartle-Hawking
vacuum, refers to a mixture of boundary conditions in the past and
future horizons. Now one requires the absence of positive energy
$U$ modes near $V=0$, and the absence of positive energy $V$ modes
near $U=0$. Physically, this mimics a black hole in thermal
equilibrium with an external heat bath. The task is to compare
these two vacuua with the Boulware vacuum which is a natural
vacuum for a fiducial observer in the asymptotic region. The
Boulware vacuum corresponds to absence of positive energy $u$ and
$v$ modes. To complete the discussion of vacua, we will also need
to take into account the reflective boundary condition at the
boundary of the anti-de Sitter space. We will do at the
end of this section.

To compute the Bogoliubov transformations between the different
natural modes, we use the standard trick \cite{Unruh:1976db}.
Let us focus on the outgoing modes first. To begin with, we
define the modes
\bea
\phi_{+,\,\omega} &=& {\rm exp}(-i \omega u) =
\theta (-U) (-U)^{\frac {i \omega}{2\pi T_H} } ~~,~~(U<0) \\
\phi_{-,\,\omega} &=& {\rm exp}(i \omega u) =
\theta (U) U^{- \frac {i \omega}{2\pi T_H} } ~~,~~(U>0) \ .
\eea
Then, to find a complete basis for positive energy
$U$ modes we consider the linear combinations which extend over
the whole $V=0$ line,
\bea
\phi_{1,\,\omega} &=& \phi_{+,\,\omega} + C_1 ~\phi_{-,\,\omega}^*
 \nonumber \\
&=& \theta (-U) (-U)^{\frac {i \omega}{2\pi T_H} }
+ C_1 ~\theta (U) U^{\frac {i \omega}{2\pi T_H}} \\
\phi_{2,\,\omega} &=& \phi_{-,\,\omega} + C_2 ~\phi_{+,\,\omega}^*
 \nonumber \\
&=& \theta (U) U^{- \frac {i \omega}{2\pi T_H} }
+ C_2 ~\theta (-U) (-U)^{- \frac {i \omega}{2\pi T_H}}
\eea
We then demand that $\phi_{1,\,\omega}$ is a positive energy
Kruskal mode: that for $\omega>0$ it must be analytic in the lower half
complex $U$-plane.  This condition is satisfied if
the coefficient $C_1$ is
\be
C_1 = {\rm exp} \left( - \frac {\omega }{2T_H} \right)
\ee
A similar condition for $\phi_{2,\omega}$ fixes the coefficient
$C_2$:
\be
C_2 = {\rm exp} \left( - \frac {\omega }{2T_H} \right)
\ee
Hence $C_1 = C_2 \equiv C$.

{}From this we can compute the (unnormalized) $\alpha$ and $\beta$
Bogoliubov coefficients which denote the overlap of a positive
energy $U$ mode with positive and negative energy $u$ modes:
\bea
\alpha = \left( \phi_{1,\,\omega}, \phi_{+,\,\omega} \right)
 = \left( \phi_{2,\,\omega}, \phi_{-,\,\omega} \right) &\propto& 1 \\
\beta  = \left( \phi_{1,\,\omega}, \phi_{-,\,\omega}^* \right)
 = \left( \phi_{2,\,\omega}, \phi_{+,\,\omega}^* \right)
&\propto& C
\eea
(for $i=1,2$). The ratio of the two coefficients is thus
\be
\left| \frac {\beta}{\alpha} \right|^2 = \left| C \right|^2 =
{\rm exp} \left( - \frac {\omega }{T_H} \right)
\label{nav}
\ee
Using the normalization condition of the Bogoliubov coefficients,
\be
\left| \alpha \right|^2 - \left| \beta \right|^2 = 1
\ee
we find that the average occupation number for
positive energy $u$ modes, seen by a fiducial observer when
the quantum field is in a vacuum with respect to positive
energy $U$ modes,
simplifies to the expected form of a Bose-Einstein
distribution,
\be
\bar{n}_{\omega} = \left| \beta \right|^2 =
\frac {1}{{\rm exp} \left( \frac {\omega}{T_H} \right) - 1 }
\ee
where $T_H$ is the Hawking temperature of the AdS$_{d+1}$ black hole.

A similar relation holds between the $V$ and $v$ modes also.
Now, we take into account the reflection from the boundary of
adS space. Ref. \cite{Klemm:1998bb} considered different
possibilities for the boundary condition at infinity: Dirichlet,
Neumann, and Robin boundary conditions. What is the
preferred boundary condition? Let us leave the geometric optics
approximation for the moment and consider mode solutions to the
exact wave equation for a free scalar field in the adS-black
hole space time. In Minkowski signature, the mode solutions
can fall into two categories, nonnormalizable
solutions $\phi^{(-)}_{\omega,\vek }$ and normalizable solutions
$\phi^{(+)}_{\omega ,\vek}$, with the asymptotic behavior
\be
   \phi^{(\pm)}_{\omega ,\vek} (t,r,\Omega ) \rightarrow
     r^{-2h_{\pm}} \tilde{\phi}^{\pm} (t,\Omega )
     \quad \quad (r\rightarrow \infty ) \ ,
\ee
where $h_{\pm}$ are parameters related to the mass $\mu = $ and the
dimension of the space $d$ by
\be
     2h_{\pm} = \frac{1}{2} (d\pm \sqrt{d^2 +4\mu^2}) \ .
\ee
The quantized field $\phi$ is expanded as a linear combination of
the normalizable modes. Their decay behavior at the boundary
corresponds to reflection. The exact mode solutions $\phi^{(+)}$ are easy to
find in $2+1$ dimensions in terms of hypergeometric
functions \cite{Keski-Vakkuri:1998nw, Ichinose:1995rg}.
Near the black hole horizon, the normalizable modes reduce to a
form
\be
    \phi^{(+)} \sim ( e^{-i\omega u} + e^{-i\omega v +
    i2\theta_0})e^{-in\phi}
\ee
where $\theta_0$ is a phase shift factor, its exact form can be
found in \cite{Keski-Vakkuri:1998nw}. Thus, in the geometric optics
approximation, the modes which take into account the
reflection from the boundary and are appropriate to a fiducial observer
are a linear combination of the positive energy $u$ and $v$ modes,
\be
  \phi_{\omega} = e^{-i\omega u} + e^{-i\omega v + i2\theta_0} \
  .
\ee
For Kruskal modes, one must consider the
corresponding linear combination of the positive energy $U$ and
$V$ modes. Thus, with the reflective boundary condition, the
appropriate vacuum is the Hartle-Hawking vacuum. Then, a fiducial observer sees
a thermal spectrum for both infalling and outgoing modes. In the
eternal adS geometry, the
Unruh vacuum is not well defined with respect to the boundary
condition at infinity. It can be viewed as an artificial
construction describing the very onset of radiation, where only outgoing
modes are thermally excited and they have not yet reflected back from the
We will return to this issue in Section 4.

\section{Hawking radiation as tunneling}

Recently, a method to describe Hawking radiation as a tunneling process,
where a particle moves in dynamical geometry, was developed by Kraus
and Wilczek \cite{Kraus:1994fh, Kraus:1995by}
and elaborated upon by Parikh and Wilczek \cite{Parikh:1999mf}. This method
also gives a leading correction to the emission rate arising from loss
of mass of the black hole correponding to the energy carried
by the radiated quantum. This method was also investigated in the
context of black holes in string theory \cite{Keski-Vakkuri:1997xp},
and it was demonstrated that
in the string picture of microstates of the black hole, the correction
to the emission rate corresponds to a difference between counting
of states in the microcanonical and canonical ensembles. However, in all
these investigations the black holes have had asymptotically flat
spacetime geometry. We now extend the investigation to black holes
in AdS spacetime. We will base our treatment on
the presentation of \cite{Parikh:1999mf}.

A convenient trick in the method of \cite{Kraus:1994fh, Kraus:1995by,
Parikh:1999mf} is to write the black hole
metric in a coordinate system where constant time slices are flat, without
a singularity at the horizon. In these coordinates, the Schwarzschild
metric takes the form
\be
  ds^2 = -(1-\frac{2M}{r})dt^2 +2\sqrt{\frac{2M}{r}} dtdr +dr^2
+r^2d\Omega^2 \ . \ee
These coordinates were first introduced 80 years ago by
Painlev\'{e} \cite{Painleve}, but then disappeared from general
knowledge, until they were independently rediscovered in
\cite{Kraus:1994fh} and used to investigate black hole quantum
mechanics. We will now derive an analogue of the Painlev\'{e}
coordinates for AdS black holes, which we shall refer to as the
AdS-Painlev\'{e} coordinates.

By analogue to the asymptotically flat black holes, the
AdS-Painlev\'{e} coordinates should have the property that
constant time slices of the AdS black hole metric (\ref{adsbh})
will have the same geometry as constant time slices of a global
AdS$_{d+1}$ metric
\be
 ds^2 = -(1+r^2)dt^2 + \frac{dr^2}{(1+r^2)} +r^2d\Omega^2_{d-1} \ .
\label{global}
\ee
Thus, we perform  a coordinate transformation $t = \that + f(r)$ so
the metric (\ref{adsbh}) takes the form
\be
ds^2 = - F(r) d\that^2 + 2 f'(r) F(r) d\that dr +
\left(  \frac {1}{F(r)} - F(r) (f'(r))^2 \right) dr^2
+ r^2 d\Omega^2
\ee
and then demand that on constant $\that$ slices the metric reduces to
\be
ds^2 = ( 1 + r^2 )^{-1} dr^2 + r^2 d\Omega^2 \ .
\ee
This implies that
\be
f'(r) = \frac {1}{r^{\frac{d-2}{2}} F(r)}
\sqrt{ \frac{\mu}{1+ r^2 }}
\ee
so the AdS-Painlev\'{e} metric reads as follows:
\be
ds^2 = - F(r) d\that^2 + \frac {2}{r^{\frac{d-2}{2}}}
\sqrt{ \frac{\mu}{1+ r^2}} d\that dr +
( 1 + r^2 )^{-1} dr^2 + r^2 d\Omega^2 \ .
\ee

Now we move on to discuss Hawking radiation. The (s-wave) quanta of
a massless scalar field follow radial light-like geodesics
\be
\dot r = - \frac {\sqrt{\mu}}{r^{\frac{d-2}{2}}} \sqrt{1+ r^2 }
\pm ( 1+  r^2 )
\label{geodesic}
\ee
where the $(+)$ sign corresponds to an outgoing geodesic and
the $(-)$ sign corresponds to an ingoing geodesic, respectively.
Next we take into account the response of the background geometry to
an emitted quantum of frequency $\omega$. We keep the total mass $M$
of the spacetime fixed, but in order to take into account
the energy carried by the quantum,
we replace $\mu$ in (\ref{geodesic}) by $\mu'$,
\be
   \mu' \equiv \frac{16\pi}{(d-1)A_{d-1}}\left( M-\omega \right) \ .
\ee
Note that at the horizon,
\be
\dot r |_{r_H} = 0 ~.
\ee
As the particle travels across the horizon from $r_{in}$ to $r_{out}$,
its action\footnote{Note that in the local point particle description used
in this section, the issue of boundary conditions at infinity does
not arise.} receives an imaginary contribution
\be
{\rm Im} ~S = {\rm Im} \int_{r_{in}}^{r_{out}} p_r ~dr =
{\rm Im} \int_{r_{in}}^{r_{out}} \int_H \frac {dH}{\dot r} ~dr
\ee
where on the last line we switched the order of integration
and used the Hamilton's equation
$\dot r =\frac{dH}{dp_r }$. Next, we substitute from (\ref{geodesic})
the radial
velocity along the outgoing geodesic, and use $dH=d(M-\omega)=-d\omega$:
\be
{\rm Im} ~S = -{\rm Im} \int_0^{\omega} d\omega'
\int_{r_{in}}^{r_{out}} dr \frac { 1+ \sqrt{\frac{\mu'}{r^{d-2}}(1+r^2)^{-1}}}
{F(r)} \ .
\ee
The only imaginary contribution to the radial integral comes from the
pole at $r_H$. Then
\be
{\rm Im} ~S = \pi \int_0^\omega d\omega' \frac {2r_H}
{r_H^2 d +(d-2)} \ .
\ee
On the other hand, after solving for $r_H$ as a function of $\mu$, we
can derive that
\be
r^{d-2}_H\frac{dr_H}{d\mu} = \frac{r_H}{r^2_H d+(d-2)} \ .
\ee
Substituting this into the integral yields
\bea
 {\rm Im}~S &=& \pi\frac{(d-1)A_{d-1}}{16\pi}
\int^{\mu}_{\mu-\frac{16\pi}{(d-1)A_{d-1}}\omega} d\mu' 2r^{d-2}_H
\frac{dr_H}{d\mu'}\nonumber \\
\mbox{} &=& \frac{1}{8} A_{d-1} \left( r^{d-1}(M)-r^{d-1}(M-\omega)\right)
 = \half \Delta S_{BH}
\eea
where $\Delta S_{BH}=S_{BH}(M)-S_{BH}(M-\omega )$ is the difference of the
entropies of the black hole before and after the emission. Thus,
the tunneling probability for the particle is
\be
  \Gamma = \exp (-\Delta S_{BH}) \ .
\label{boltz}
\ee
If we Taylor expand $\Delta S_{BH}$ in $\omega$, the leading term
gives the thermal Boltzmann factors  $\exp(-\omega /T_H)$
for the emanating radiation. The second term represents
corrections from the response of the background geometry to the
emission of a quantum.
The same result holds for emission from asymptotically
flat black holes \cite{Keski-Vakkuri:1997xp}.

\section{Particle creation by a collapsing spherical shell in AdS}

We will now turn to a third way to analyze Hawking radiation from
black holes, and investigate particle creation by a collapsing
spherical body which forms a black hole in AdS. We will base our
treatment on the discussion in \cite{Birrell:1982ix},
which in turn follows \cite{Hawking:1974sw} and
\cite{Parker:1975jm}.
As in \cite{Birrell:1982ix}, the starting point is that we assume that in
the remote past the spherical body is distended so much that it
deforms the anti-de Sitter space. Thus, in the beginning we can assume
that a quantum field is in a vacuum constructed with respect to global
coordinates in AdS space. Now a convenient choice for the global
coordinates is given by
\be
   ds^2 = (\sec \rho )^2 (-dt^2+d\rho^2)+(\tan \rho )^2 d\Omega^2_{d-1} \ .
\label{global2}
\ee
The (normalizable)
mode solutions can be found e.g. in \cite{Balasub:1998sn}:
\be
  \phi^{(+)}_{n,l} = e^{-i\omega t}Y_{l,\{m\}}(\Omega )
 (\cos \rho )^{2h_+} (\sin \rho )^l
 P^{(l+\frac{d}{2}-1,2h_+-\frac{d}{2})}_n (\cos 2\rho ) \ ,
\label{glomode}
\ee
where $Y_{l,\{m\}}(\Omega )$ is a spherical harmonic on $S^{d-1}$,
$P^{(l+\frac{d}{2}-1,2h_+-\frac{d}{2})}_n$ is a Jacobi
polynomial, and
\be
   2h_+ = \frac{d}{2}+\half\sqrt{d^2+4m^2} \equiv \frac{d}{2} +\half \nu
\ee
Due to the boundary conditions at the origin and at the boundary of AdS,
the spectrum is discrete, with
\be
  \omega = 2h_+ +2n +2l \ ; \ n=0,1,2,\ldots
\ee
Again, we will focus on $s$-waves ($l=0$). As in flat space, we expect
the quantum to experience a strong redshift as it propagates across the
collapsing body. Thus, in the remote past, we are most interested in
high frequency modes. In the high frequency limit,
the mode solution (\ref{glomode})
reduces to a simplified form
\be
  \phi^{(+)}_n \sim (\cos \rho )^{(d-1)/2}
e^{-i\omega_nt} \cos(\omega_n \rho -\frac{d\pi}{4} )
\label{glosimp}
\ee
where we suppressed normalization factors. In other words, they take
a form of a standing wave, a superposition of an ingoing and outgoing
spherical wave, with a discrete spectrum. Note that using the radial
coordinate $r$, the overall
factor $(\cos \rho )^{(d-1)/2} \sim (1/r)^{(d-1)/2}$
as $r >>1$, so we recover the expected overall decay factor for the amplitude.
Now, we will add into the
picture the collapsing body and try to compute the redshift due
to the passage of the wave across it. As in \cite{Birrell:1982ix},
we will use the
geometric optics approximation, and truncate the
analysis to two dimensions to the $t,r$-plane by suppressing
the overall decay factor of the waveform. For simplicity, we
assume that the collapsing body is a thin shell of radius $R$, with $R$
monotonically decreasing in time. The truncated metric inside and outside
the shell takes the form
\be
    ds^2 = -F_{\pm}dt_{\pm}^2 +F_{\pm}^{-1}(r)dr^2
\ee
where $F_+(r)=1-\frac{\mu}{r^{d-2}}+r^2$ and $F_-(r)=1+r^2$. We then define
the tortoise coordinates $r^{\pm}_*$,
\be
    r^{\pm}_* = \int \frac{dr}{F_{\pm}(r)} \ ,
\ee
and the null coordinates
\bea
   u &=& t_+ - r^+_* \ , \  v = t_+ + r^+_* \nonumber \\
   U &=& t_- - r^-_* \ , \  V = t_- + r^-_*
\eea
so that the interior and exterior metrics are conformal to a flat metric.
The tortoise coordinate in the interior is  $r^-_*=\arctan r$,
so the origin $r=0$ corresponds to $r^-_*=0$.
In terms of the null
coordinates $U,V$, the origin is then at $V-U=0$. Note also that
$r^-_*=\rho$, where $\rho$ is the coordinate that appears in the
global metric (\ref{global2}).
The exterior and
interior null coordinates are related by
\bea
    v &=& \beta (V) \nonumber \\
    U &=& \alpha (u)
\eea
where $\alpha (u)$ and $\beta (V)$ are to be determined below. In the
$(t,r)$ coordinates, the passage of a wave across the shell turns to
a reflection condition at the origin:
\be
  v = \beta (V) = \beta (U) = \beta (\alpha (u)) \ .
\ee
Thus, in the asymptotic region (near the boundary), the waves have
a phase structure
\be
 \tilde{\phi}^{(+)} \sim e^{-i\omega_n v}-e^{-i\omega_n \beta (\alpha (u))} \ .
\ee
To find the functions $\alpha ,\beta$, we match the interior and exterior
metrics across the collapsing shell at $r=R(\tau )$. Here $\tau$ denotes
the shell time, which is related to the time coordinates $t_{\pm}$ in the
interior and exterior of the shell through
\be
  ds^2 = [-F_{\pm}dt^2_{\pm}+F^{-1}_{\pm}dr^2]_{|r=R(\tau )}
       = -d\tau^2 \ .
\label{stime}
\ee
It is easiest to consider
the derivatives
\bea
   \alpha'(u) &=& \frac{dU}{du} = \frac{\Udot}{\udot}\nonumber \\
   \beta'(V) &=& \frac{dv}{dV} = \frac{\vdot}{\Vdot}\ ,
\eea
(at the shell) where $\cdot = d/d\tau$. Using the definition (\ref{stime}),
we obtain
\bea
   \frac{dU}{du} &=& \frac{F_+(R)[\sqrt{F_-(R)+\Rdot^2}-\Rdot ]}
                     {F_-(R)[\sqrt{F_+(R)+\Rdot^2}-\Rdot ]} \nonumber \\
   \frac{dv}{dV} &=& \frac{F_-(R)[\sqrt{F_+(R)+\Rdot^2}+\Rdot ]}
                     {F_+(R)[\sqrt{F_-(R)+\Rdot^2}+\Rdot ]}  \ .
\eea
As the radius of the shell approaches the horizon, we can approximate
\bea
  F_+(R) &\approx& 4\pi T_H(R-r_H) \nonumber \\
  F_-(R) &\approx& F_-(r_H) \equiv A \ .
\eea
Then, we can approximate
\be
  \frac{dU}{du} \approx -\frac{2\pi T_H(R-r_H)}{A\Rdot}
 [\sqrt{A+\Rdot^2}-\Rdot ]
\ee
where we used $\sqrt{\Rdot^2}=-\Rdot$ since $\Rdot<0$ as the shell is
collapsing and $|\Rdot|\neq 0$ as a function of shell time. In the above,
$\Rdot = \Rdot_{|r_H}$. Next, we relate $U$ to $R-r_H$ by expanding
\be
  U \approx U(r_H)+\frac{dU}{dR}_{|R=r_H} (R-r_H)
\ee
and evaluating the derivative $dU/dR$ at the horizon, using the chain
rule and the definitions () and (). We obtain
\be
  (R-r_H) \approx (U-U(r_H))\frac{A\Rdot}{[\sqrt{A+\Rdot^2}-\Rdot]} \ .
\ee
We substitute this to (), and obtain
\be
   \frac{dU}{du} \approx -2\pi T_H (U-U(r_H)) \equiv -\kappa U + const.
\ee
where $\kappa$ is the surface gravity of the black hole. Integration
then gives
\be
   \alpha (u) = e^{-\kappa u} + const.
\ee
A similar calculation for $dv/dV$ gives
\be
  \frac{dv}{dV} \approx -\frac{A}{\Rdot [\sqrt{A+\Rdot^2}+\Rdot ]}
   \equiv c  \quad \quad (=const.) \ .
\ee
By integration,
\be
   \beta (V) = cV+const.
\ee
Thus, we find that in the asymptotic region
the waves have a phase structure
\be
 \tilde{\phi}^{(+)}
 \sim e^{-i\omega_n v} - e^{-i\omega_n c(e^{-\kappa u}+const.)} \ .
\ee
To obtain modes where the outgoing wave is of standard form, we
invert functionally and write
\be
\tilde{\phi}^{(+)}
 \sim e^{i\omega_n \kappa^{-1}\ln [(v_0-v)/c]} -e^{-i\omega_n u} \ ,
\ee
which is valid only for $v<v_0$. Now, we move back to $(d+1)$
dimensions and compare with the high-frequency limit of the global modes
(\ref{glosimp}).
Note that far in the asymptotic region, $r\rightarrow \infty$, the
exterior tortoise coordinate reduces to
the same form as $\rho$, $r^+_* \approx \rho$, so we can write the
global mode as
\be
   \phi^{(+)} \sim (\cos \rho )^{(d-2)/2}
 (e^{-i\omega_n v-id\pi/2}-e^{-i\omega_n u}) \ .
\ee
We want to compare this with the modes that we found in the collapsing
shell geometry,
\be
  \tilde{\phi}^{(+)} \sim (\cos \rho )^{(d-2)/2}
 (e^{i\omega_n \kappa^{-1} \ln [(v_0-v)/c]} -e^{-i\omega_n u} ) \ \ (v<v_0) \ ,
\ee
where we added the overall decay factor. The Bogoliubov transformation
follows the discussion in \cite{Birrell:1982ix}, and as a result
we find that the outgoing modes are thermally excited, if the
field is in a global vacuum. Thus, the global vacuum resembles
an Unruh vacuum.

Naturally, the above result only applies to the onset of Hawking
radiation as the black hole has formed, and does not address the
issue of subsequent evolution.
Since the AdS space can be viewed as a finite
box, what will happen is that the outgoing radiation cannot escape
to infinity but will slowly fill the box. Subsequently, the black
hole will come to a thermal equilibrium with the surrounding thermal
bath. This situation is descibed by the Hartle-Hawking vacuum
in the eternal geometry. Note however that
in dimensions $d>2$,
if the initial size of the black hole is too small,
the equilibrium will not be established before the black hole evaporates
completely. See \cite{Horowitz:1999jd} for a thorough analysis.

\bigskip

\noindent
{\large {\bf Note added}}

\bigskip

As we were finalizing this paper, the paper \cite{SAA} appeared,
discussing Hawking radiation in the optical collapse geometry
for spherically symmetric black holes.

\bigskip

\noindent
{\large {\bf Acknowledgment}}

\bigskip

We would like to thank Jorma Louko for useful comments.

\end{document}